# Measurement of the Velocity of the Neutrino with MINOS

P. Adamson[1†], N. Ashby[2], R. Bumgarner[3], M. Christensen[3], B. Fonville[3], J. Hirschauer[3]
S. R. Jefferts[2], D. Matsakis[3], A. McKinley[3], T. E. Parker[2], E. Powers[3], S. Römisch[2],
J. Wright[3], and V. Zhang[2]

[1]Fermi National Accelerator Laboratory, Batavia, Illinois 60510, USA

[2]NIST Time and Frequency Division, Boulder, Colorado 80305, USA

[3]USNO, Washington, DC 20392, USA

pa@fnal.gov

**Abstract—The MINOS experiment uses a beam of predominantly muon-type neutrinos generated using protons from the Main Injector at Fermilab in Batavia, IL, and travelling 735 km through the Earth to a disused iron mine in Soudan, MN. The 10 μs-long beam pulse contains fine time structure which allows a precise measurement of the neutrino time of flight to be made. The time structure of the parent proton pulse is measured in the beamline after extraction from the Main Injector, and neutrino interactions are timestamped at the Fermilab site in the Near Detector (ND), and at the Soudan site in the Far Detector (FD).**

**Small, transportable auxiliary detectors, consisting of scintillator planes and associated readout electronics, are used to measure the relative latency between the two large detectors. Time at each location is measured with respect to HP5071A Cesium clocks, and time is transferred using GPS Precise Point Positioning (PPP) solutions for the clock offset at each location.**

**We describe the timing calibration of the detectors and derive a measurement of the neutrino velocity, based on data from March and April 2012. We discuss the prospects for further improvements that would yield a still more accurate result.**

## I. Background

The earliest measurements of the speed of the neutrino were made in the 1970s, with the Fermilab Main Ring narrow-band neutrino beam. By comparing the arrival times of neutrinos and muons, a limit on the fractional difference between the neutrino speed and the speed of light $|v/c - 1| < 4 \times 10^{-5}$ is obtained at the 95% confidence level [1] with neutrinos of energies 30-200 GeV over a baseline of about 900 m.

The observation, by the Kamiokande [2], IMB [3], and Baksan [4] experiments of 24 neutrino events from supernova SN1987A within a 30 s period allows one to place a tighter constraint on the velocity of these neutrinos. The first light from SN1987A was observed some 3 hours after the neutrino pulse, but as nobody was watching the Large Magellanic Cloud in the hours prior to this observation, this is only a limit. Combining this observation with supernova models, Longo [5] derives a limit $|v/c - 1| < 2 \times 10^{-9}$ for the speed of neutrinos with energies 10-40 MeV, over a baseline of 168,000 light years.

Tighter constraints are derived by Cohen and Glashow [6], arguing that superluminal neutrinos must produce $e^+/e^-$ pairs via brehmsstrahlung, and by Davies and Moss [7] with arguments based on nucleosynthesis in the early universe and acoustic oscillations in the cosmic microwave background, but these are not direct measurements of the neutrino velocity.

† For the MINOS collaboration.

In September 2011, the OPERA experiment reported an observation of neutrinos from CERN arriving 730 km away at the Gran Sasso laboratory 60 ns earlier than the speed of light would predict [8], corresponding to neutrinos travelling faster than light by $(v/c - 1) = (2.37 \pm 0.45) \times 10^{-5}$. This measurement was subsequently found to be in error; the OPERA collaboration reported their amended result, which is completely consistent with neutrinos travelling at the speed of light, in [9].

In response to the September 2011 claims, the MINOS experiment, in collaboration with NIST and USNO, installed an upgraded timing system with the aim of measuring the neutrino time of flight with 1 ns accuracy. This paper reports the initial results of this effort.

## II. OVERVIEW

MINOS [10] is a long baseline neutrino oscillation experiment. Protons from the Main Injector accelerator at Fermilab incident on a graphite target produce a beam of predominantly muon neutrinos. The neutrino beam is measured at the Near Detector (ND) 1.04 km from the target, on the Fermilab site, then passes through the Earth under Wisconsin and is measured again 735 km from the target, at the Far Detector (FD) in Soudan, MN. The oscillation measurement [11] compares the flavor content of the neutrino beam at the two detectors and determines the mass and mixing parameters.

The neutrino velocity measurement is conceptually straightforward, consisting of a measurement of the distance between two detectors and the time it takes for a neutrino to pass between them. However, the interaction cross-section for the neutrino is very small (meaning that most neutrinos travel through the earth, both detectors, and head off into space without leaving a trace of their presence) and the process of detecting a neutrino is destructive. This means that we can never observe the same neutrinos in both detectors, but take a small, independent random sample of the neutrino beam in each one. It is therefore necessary to consider the time distribution of the neutrino beam.

### A. *Time structure of neutrino beam*

The neutrinos are produced in the decays of pi and K mesons in the decay pipe of the NuMI (Neutrinos at the Main Injector) facility at Fermilab, and these mesons in turn are formed in the interaction of the primary 120 GeV proton beam with a graphite target, as shown in Figure 1. The mesons that produce our neutrinos have a range of energies mostly between abut 5 and 15 GeV, and so have a range of velocities. In addition, their decay is a random process, so each “slow” meson travels a different distance before decaying to a fast neutrino. The combination of these two effects introduces a random jitter of about 100 ps on the time between the parent proton striking the target and the neutrino passing a fixed point at the end of the facility.

The parent 120 GeV protons have a time distribution inherited from the 53 MHz RF structure of the Main Injector accelerator. Protons destined for the NuMI facility are grouped in six “batches,” each approximately 1.6 μs long, separated by about 100 ns. Each batch consists of 81 bunches, each separated by 18.83 ns (the RF frequency of the Main Injector at 120 GeV is 53.103480 MHz), and each bunch has a sigma of roughly 1ns, and a full width of 3.5 ns (see Figure 2).

### B. *Detectors*

The two MINOS detectors are functionally equivalent magnetized steel/scintillator tracking calorimeters. They are longitudinally segmented, consisting of alternating planes of 1” thick steel and 1 cm thick plastic scintillator. The scintillator planes consist of long, 4.1 cm wide strips of plastic scintillator, read out with wavelength-shifting fibers and multi-anode photomultiplier tubes, and are alternately oriented at ±45° with respect to the horizontal. The Near Detector weighs 0.98 kt, and the Far Detector 5.4 kt.

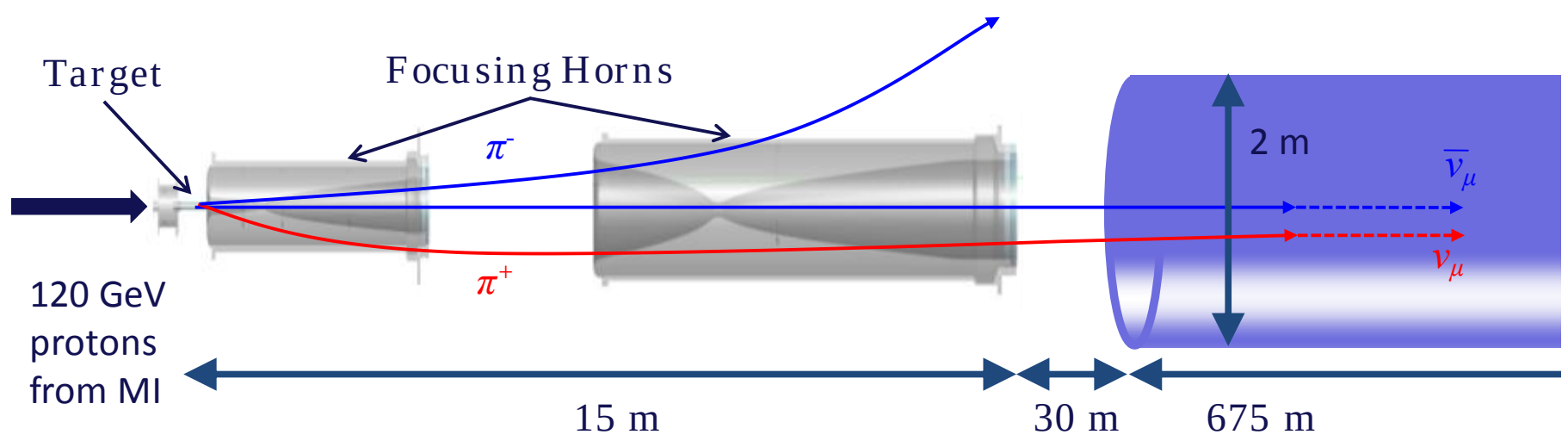

Figure 1. Neutrino production in the NuMI facility.

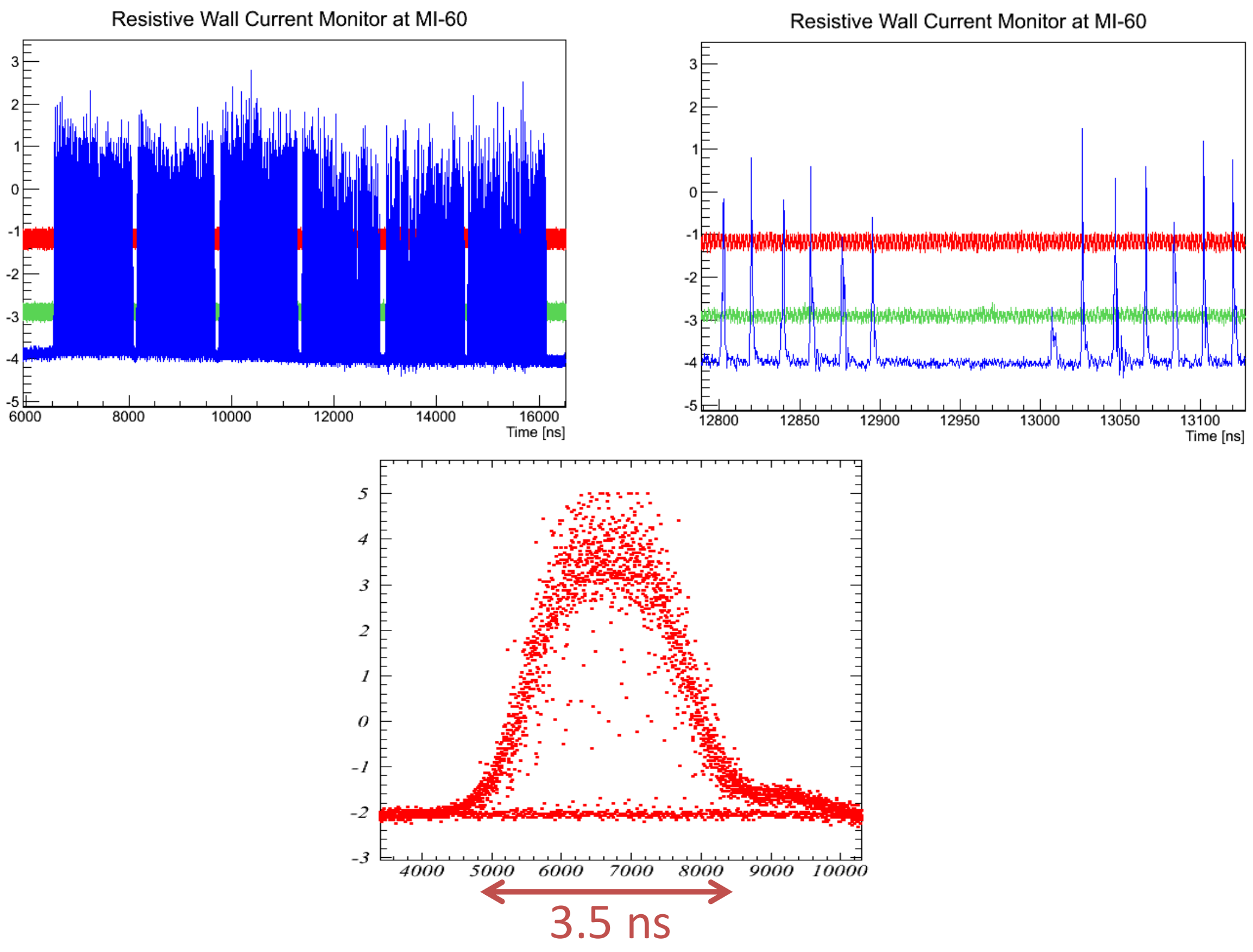

Figure 2. Time structure of NuMI protons, showing the whole spill (top left), the gap between two batches (top right) and a single bunch (bottom).

Muon neutrino charged-current interactions (the only interactions considered in this analysis) produce a long muon track that passes through tens of scintillator planes. Each scintillator strip hit by a muon results in a timestamped hit in an electronics channel. Fitting this set of hits under the constraint that the muon travels at the speed of light (which is close enough to true over the short distances involved) results in a position and time ($x,y,z,t$) for the interaction of the neutrino in the detector and the generation of the muon. This measurement is observed to have 1.5 ns resolution.

*C. Old Timing System*

The time of recorded neutrino interactions in the MINOS detectors is used to reject background from muons from cosmic rays. Each detector has a Truetime‡ XL-AK single-frequency GPS timing receiver, which is used to derive an absolute timestamp for the neutrino events. A neutrino pulse is about 10 µs long, and occurs every 2 s or so, so requiring a neutrino event at the Far Detector to be in time with the beam pulse reduces the background by a factor of more than 200,000. The complete MINOS dataset, from seven years of running, contains about 4000 charged-current $\nu_\mu$ events due to the neutrino beam and a single background event. Such background rejection does not need the event time to be accurate to much better than a microsecond, and the Truetime receivers are more than adequate for this need. With this system and seven years of data, we achieved a measurement of the difference between the time of flight of the neutrino over 734 km and the expectation assuming travel at the speed of light of $\delta$ = -15 ± 31 ns. To do better required an improved timing system.

## III. NEW TIMING SYSTEM

A critical requirement of the new system was to not interrupt the MINOS experiment data-taking. To this end, we left the old system undisturbed, and implemented a new precision timing system to measure what the old system was actually doing. An overview of the new system is shown in Figure 3.

We consider three locations. The first is the MI-60 building, an accelerator service building situated at the point where the NuMI beamline leaves the Main Injector. Here, in the NuMI beamline, is a Resistive Wall Current Monitor, a device that measures the time distribution of the proton beam by measuring the current induced in the beam pipe. This particular device has 1 GHz analog bandwidth, and the signal runs from the tunnel to the surface building via 161.1 ns of Andrew Heliax LDF-6.

The second location is the Near Detector building, about 1.4 km away, on the surface about 100 m above the Near Detector, and the third is the Far Detector, with a building on the surface about 700 m above the detector. In these two neutrino detector locations, all the electronics resides underground next to the detectors.

Each location has an atomic frequency standard, against which various detector electronics signals are measured. At MI60, this is an SRS FS725 rubidium standard, and at the Near and Far detectors it is a standard performance HP5071A. The philosophy is to measure the detectors against the atomic clock, then transport the clock signals to the surface with two-way time transfer, and measure the clocks with modern dual-frequency GPS receivers.

‡ Commercial products used in this experiment are named for clarity and completeness. We do not endorse any particular product, and make no claim that our measurements are typical of the products quoted.

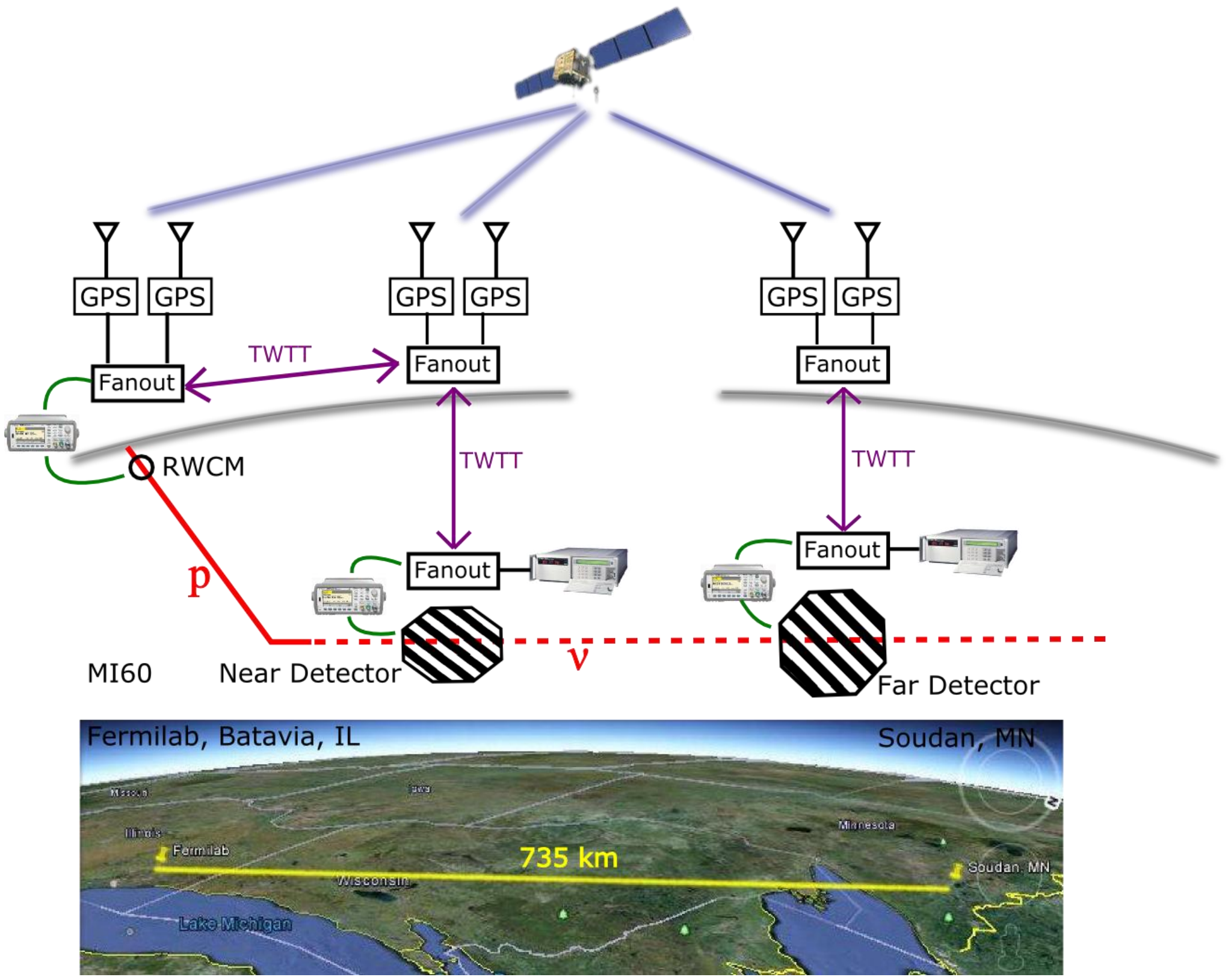

Figure 3. New timing system overview.

## A. MI60

At MI60, the Resistive Wall Current Monitor (RWCM) signal is digitized by a Tektronix TDS7154 oscilloscope at 2.5 GS/s. The 'scope is referenced to the 10 MHz signal from the local Rb clock, and is triggered by a beam-synchronous signal derived from the accelerator control system that is coincident with beam extraction from the Main Injector. The proton signal waveform and trigger are recorded, and the time of the trigger with respect to the Rb clock pulse per second (PPS) signal is measured with an Agilent 53230A Time Interval Counter.

## B. Near Detector

The ND is triggered by a signal known as SGATE, which is derived from the beam-synchronous accelerator clock signal mentioned above. An additional diagnostic electronics board was installed into the ND electronics in order to extract the SGATE signal and measure it against the local Cs clock PPS with an Agilent 53230A. The ND electronics records the time between SGATE and the neutrino event with respect to the ND 53.1 MHz clock, which is sourced from a VCXO with the voltage control held at constant voltage. Unfortunately, this clock is imperfectly isolated from the Main Injector machine clock, and shows a 100 Hz variation synchronous with the accelerator cycle. As we only use this timebase to measure the time between SGATE and the neutrino event (a time of order 20 μs), the effect of the clock variation is roughly 40 ps at worst, and can be neglected.

## C. Far Detector

The FD clock is derived from the Truetime GPS receiver. The 10 MHz GPS clock is multiplied to make 40 MHz, which is distributed to each of 16 VME readout crates. It is distributed over a custom backplane to each readout board in the crate, where it is multiplied again to produce four quadrature copies of an 80 MHz clock. The timing signal from the detector is split, and the use of a 1.56 ns delay line produces the final bit for an effective 640 MHz clock. The detector PPS is measured against the local Cs clock with an Agilent 53230A, and is observed to vary by typically ±10 ns in the short term. The frequency of the backplane 40 MHz is also recorded, and can be used to correct the detector timestamps.

The relative timing of the multiple VME crates in each detector is verified with through-going cosmic ray muons. These produce straight, easily-measured tracks that should travel at the speed of light, and we verify that there is no step change in time at a crate boundary.

*D. Auxiliary Detectors*

The systems above allow us to time the detectors against a local atomic clock, but there is an outstanding unknown latency in each neutrino detector. In principle, one could measure the delay of the various paths through the complicated detector electronics and arrive at an answer, but such an attempt is prone to error. We choose instead to measure the latency of the whole neutrino detector and electronics system in situ.

We build two identical, small, portable "auxiliary detectors (ADs)," each consisting of a pair of 50 cm × 50 cm planes of MINOS-style plastic scintillator read out by the same Hamamatsu M16 multi-anode phototubes used by the MINOS Far Detector. A coincidence between the PMT dynode signal from each of the two planes in a detector signals a muon passing through the detector. We place both detectors behind the MINOS Near Detector, and record the time of muons in the detector with respect to the local Cs clock PPS with a Brilliant Instrument BI220 Time Interval Analyser. Matching muon hits in time with muons recorded in the Near Detector, expressed with respect to the Cs clock with the equipment described in section B above and correcting for the longitudinal position of the auxiliary detector, we obtain a relative latency measurement between the two ADs and between the ADs and the Near Detector.

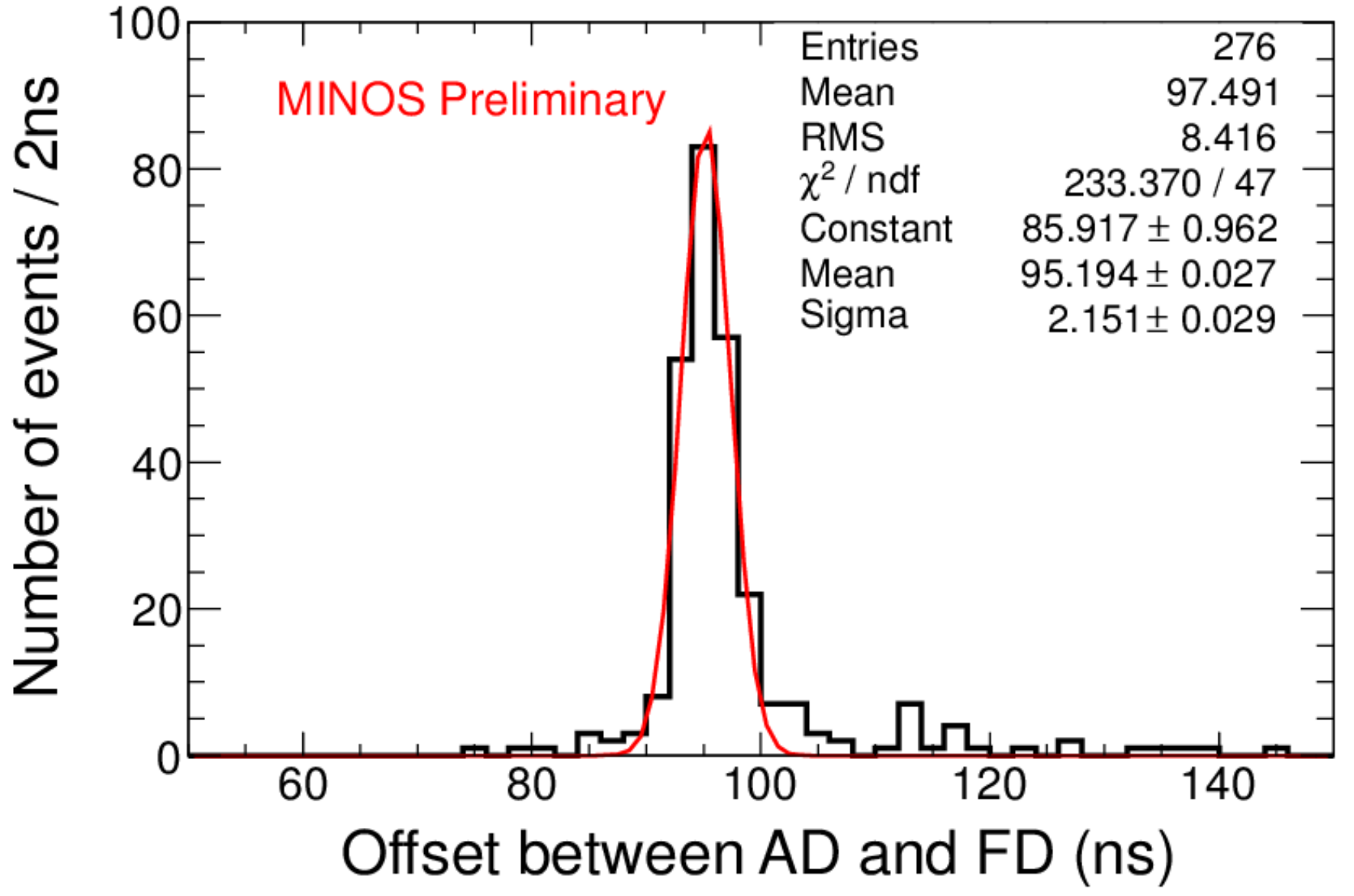

Figure 4. Measuring Far Detector latency with the auxiliary detector.

One complete detector and readout electronics is then transported to the Far Detector, where the measurement is repeated. We now have a measurement of the relative latency of the ND and FD, which is exactly what we need for a time-of-flight measurement. Figure 4 shows the measurement of the relative latency at the FD, up to some constant cable offsets. The tail of the distribution is caused by noise in the coincidence unit. We measure a relative latency of 36 ns at the ND and 12 ns at the FD, so have a relative latency of 24 ± 1 ns. The error assigned covers both the uncertainty on the initial relative latency measurement and small drifts observed in the measurement over time. Ideally, one would return the traveling AD to the Near Detector in order to close the measurement. We will be able to do this later in 2013, when the accelerator upgrades are completed and the beam returns. Given that the AD has fairly straightforward electronics, we can also estimate the absolute latency of each detector. Because the delays in the AD itself don't cancel in an absolute measurement, this has an uncertainty of 4 ns. This error will not feature in the final time of flight measurement. Note that our definition of latency also encompasses any errors in the delay measurements for Sections B and C above.

*E. Two-way time transfer to surface*

The neutrino detectors are 100m (ND) and 700m (FD) underground. PPS and 10 MHz signals from the underground Cs clock fanouts are sent to the surface over single-mode fiber with Linear Photonics Timelink modules. At the surface, these signals are replicated in a second set of fanouts, and then sent back underground over another single-mode fiber from the same bundle. The round-trip time of the PPS signal is measured underground with a TIC.

The links are calibrated by exchanging the upward and downward-going fibers, and observing the shift in the upstairs fanout with respect to an additional clock. We exchange Timelink modules for a spare set to measure the internal delays of each of the modules – we find in our units that the transmitters all have the same delay within 100 ps, but the receiver delays vary by up to 2 ns. We measure the internal delay through the fanouts.

As a cross-check, we performed a "clock trip" measurement to verify the calibration of our links. The offset between the surface PPS fanout and a standard-performance HP5071A was measured with an SR620 TIC. The clock and TIC were carried underground, where the offset with respect to the underground fanout was recorded. Returning the clock to the surface, we re-measured the offset upstairs, so estimating and correcting for the relative drift of the stationary and traveling Cs clocks during the measurement. This measurement at the ND is shown in Figure 5, showing an average discrepancy of 400 ps. At the Far Detector, we did not have enough spares available to measure the delay of the Linear Photonics receivers. We did make three clock trip measurements on different days, and so use the mean of these measurements to correct the time between upstairs and downstairs, and apply the ±0.55 ns range as an additional systematic uncertainty.

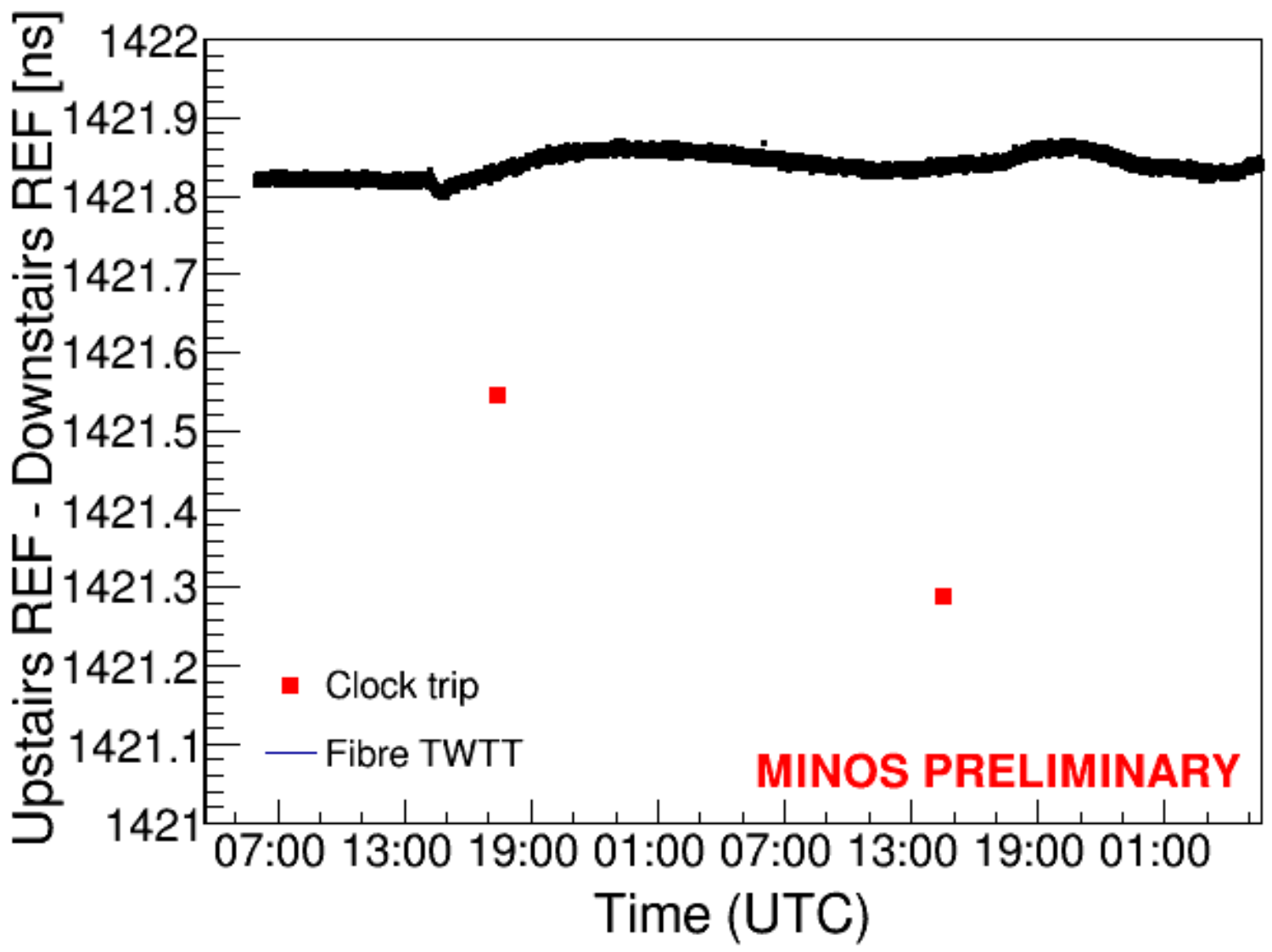

Figure 5. Clock trip check of TWTT at Near Detector.

*F. GPS systems*

We use 8 similar, modern dual-frequency GPS receivers in the experiment. GPS1-6 are NovAtel OEMV receivers with 703-GG antennas. GPS 7 and 8 are newer NovAtel OEM6, with 702-GG antennas. All units use temperature-annealed Andrew FSJ-1 as antenna cables – the shortest is about 19 m and the longest about 40 m. GPS 1 and 7 are located at MI60, GPS 2 and 8 at the ND on the Fermilab site, and GPS 3 and 4 at the FD in Soudan, MN.

GPS 5 and 6 are traveling receivers, rotated around the three MINOS sites and NIST in Boulder, CO, in order to cross-calibrate the fixed GPS systems and eliminate the possibility of the internal delay of the receivers at one site drifting with respect to another set. The traveling GPS travel as a bundle with their antenna cable and antenna. The GPS systems are described in more detail in [12].

All GPS data used in this paper are PPP solutions obtained using the online NRCan software. PPP solutions were computed for 5-day periods, and the center day's data used each time. We exclude some periods featuring known problems, such as the day the high Chicagoland winds blew a traveling GPS antenna on its side. Figure 6 shows data from MI60 and the FD. We suspect that the better stability shown by the FD units is due to their being placed into a controlled environment that maintains the temperature to within 1°F, whereas the ND and MI60 units are in ordinary indoor room environments. We derive a systematic uncertainty of 0.5 ns on the GPS timing based on the repeatability of the traveling GPS measurements. This is a little more optimistic than the estimate in [12], and is caused by a difference in the selection of the data. This uncertainty does not contribute significantly to the final result, so we do not pursue this difference further, trusting to more time and data to clarify the situation in the future.

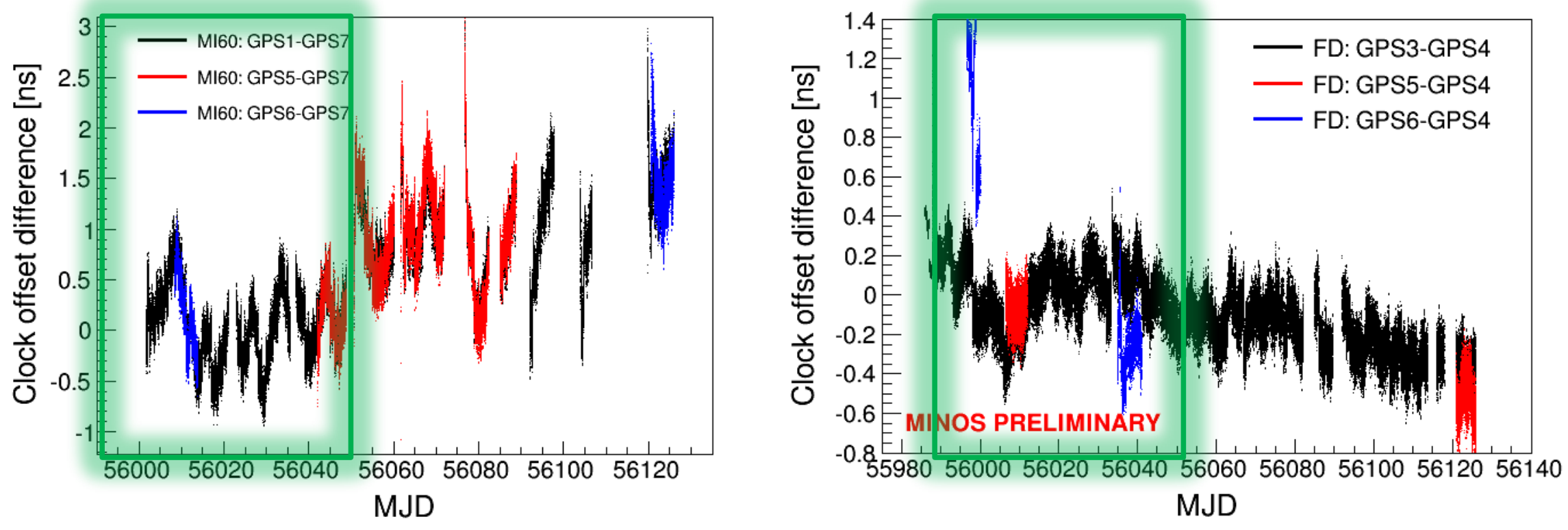

Figure 6. Differences between two GPS units (common clock) at MI60 (left) and FD (right). The green shaded area shows the data used in this analysis. The black points are fixed GPSs and the red and blue the difference between a travelling GPS and a fixed one.

We cross-checked the GPS calibrations with a TWSTFT link measurement between ND and FD, made by USNO personnel on April 18th-19th 2012. This comparison is shown in Figure 7, and shows agreement between TWSTFT and GPS at the 500 ps level.

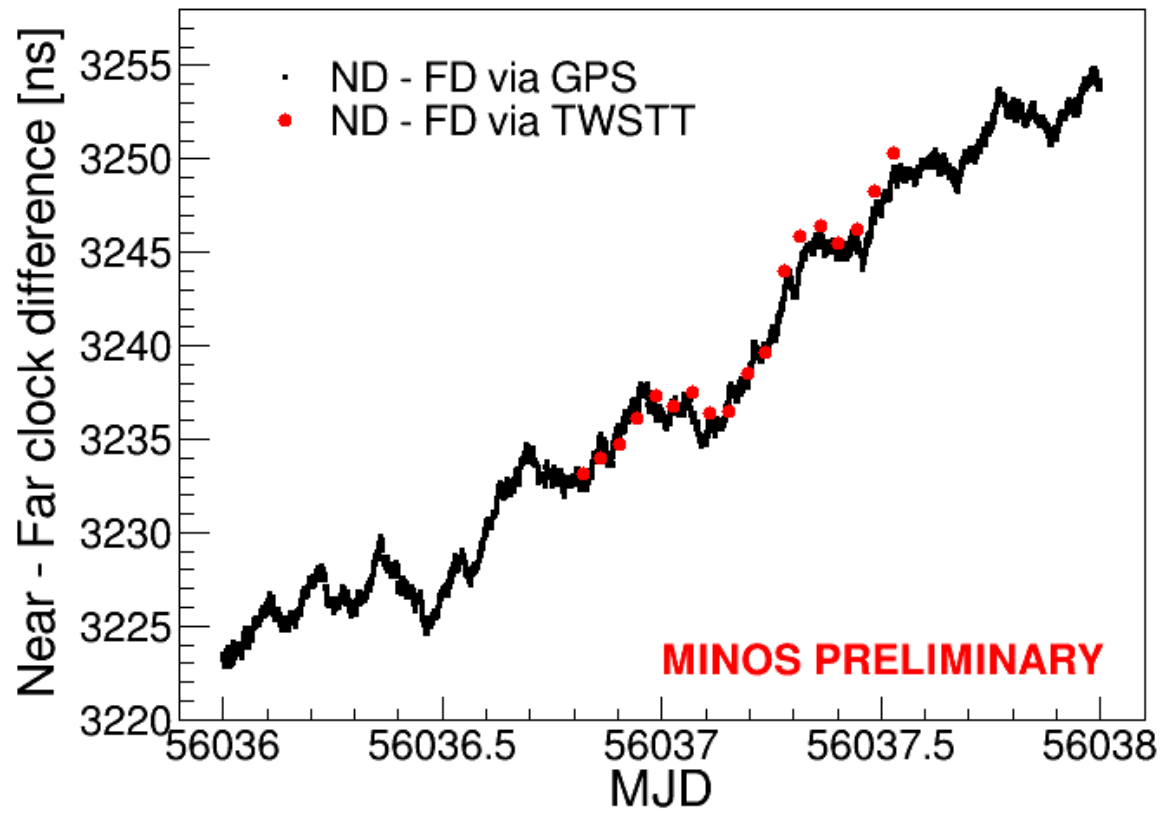

Figure 7. Offset between ND and FD surface PPS signals, measured with GPS and TWSTT.

## IV. Analysis

The analysis of the data in this experiment was performed blind, in order to eliminate the possibility of someone's subjective bias about the speed of neutrinos affecting the result (either subconsciously or not). The analysis methodology and the systematic errors were finalized before looking at the measurement of the distance between the detectors.

### *A. Methodology*

For each neutrino event recorded in the detector, we find the associated distribution of protons as seen by the MI60 RWCM. As we do not know a priori which proton bunch produced our neutrino, we have a 10 μs-wide range of possible values for the neutrino time of flight, and form a likelihood function from the proton waveform by convolving it with a 1.5 ns sigma Gaussian distribution representing the timing resolution of the neutrino detectors.

As we acquire more neutrino events, we multiply the likelihood functions to obtain an overall likelihood for the dataset. As shown in Figure 8, as we acquire more data, we rule out more and more possibilities for the time of flight until we are left with a single narrow peak.

### *B. MI60 to ND crosscheck*

Before we un-blind the measurements involving the Far Detector, we can check that everything is working by measuring the time of flight between MI60 and the Near Detector by the same method. Because of the high statistics available at the ND, we can measure the time of flight on a daily basis over the run period. We know the distance between the MI60 RWCM and the ND along the beamline [13], so can calculate the expected time of flight, and compare to our measurement.

As another cross-check, we have a TWTT link over fiber between MI60 and ND, using the same hardware as the links between upstairs and underground at the two neutrino detectors. This means we can use either GPS or fiber TWTT to do the time transfer between MI60 and ND, and can calculate the time of flight each way. We show the results from this measurement in Figure 9. From the surveyed distance between the MI60 RWCM and the ND [13], and the absolute latency of the ND measured with the Auxiliary Detector, we expect the time of flight to be 4622.7 ± 4.0 ns, where the uncertainty comes from the absolute latency of the auxiliary detector. As we see in Figure 9, our measurement is comfortably consistent with this. The figure also gives us an alternative measurement of the stability of the time transfer – the time of flight of the proton / pion / neutrino beam should be constant, so given that we see that the TWTT is stable at better than 50 ps, whereas the GPS time transfer shows a 200 ps systematic variation.

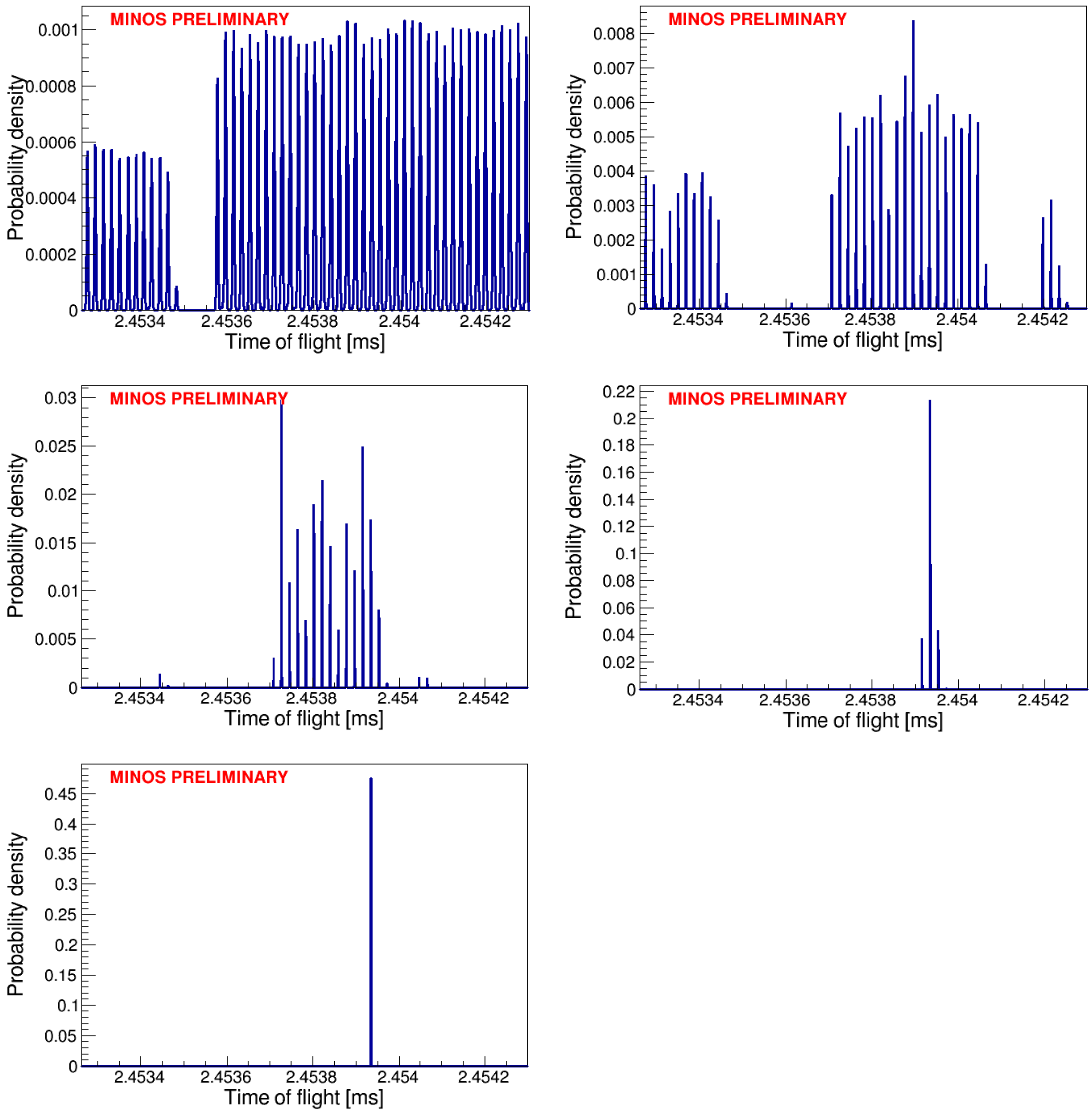

Figure 8. Portion of likelihood function for neutrino time of flight between MI60 and Far Detector for 1 (top left), 10 (top right), 30 (center left), 60 (center right) and 195 (bottom left) events. As we acquire more data, large ranges of time of flight are excluded because of the coarse time structure of the neutrino spill, and each individual peak narrows as the phase of the 53 MHz structure is better measured.

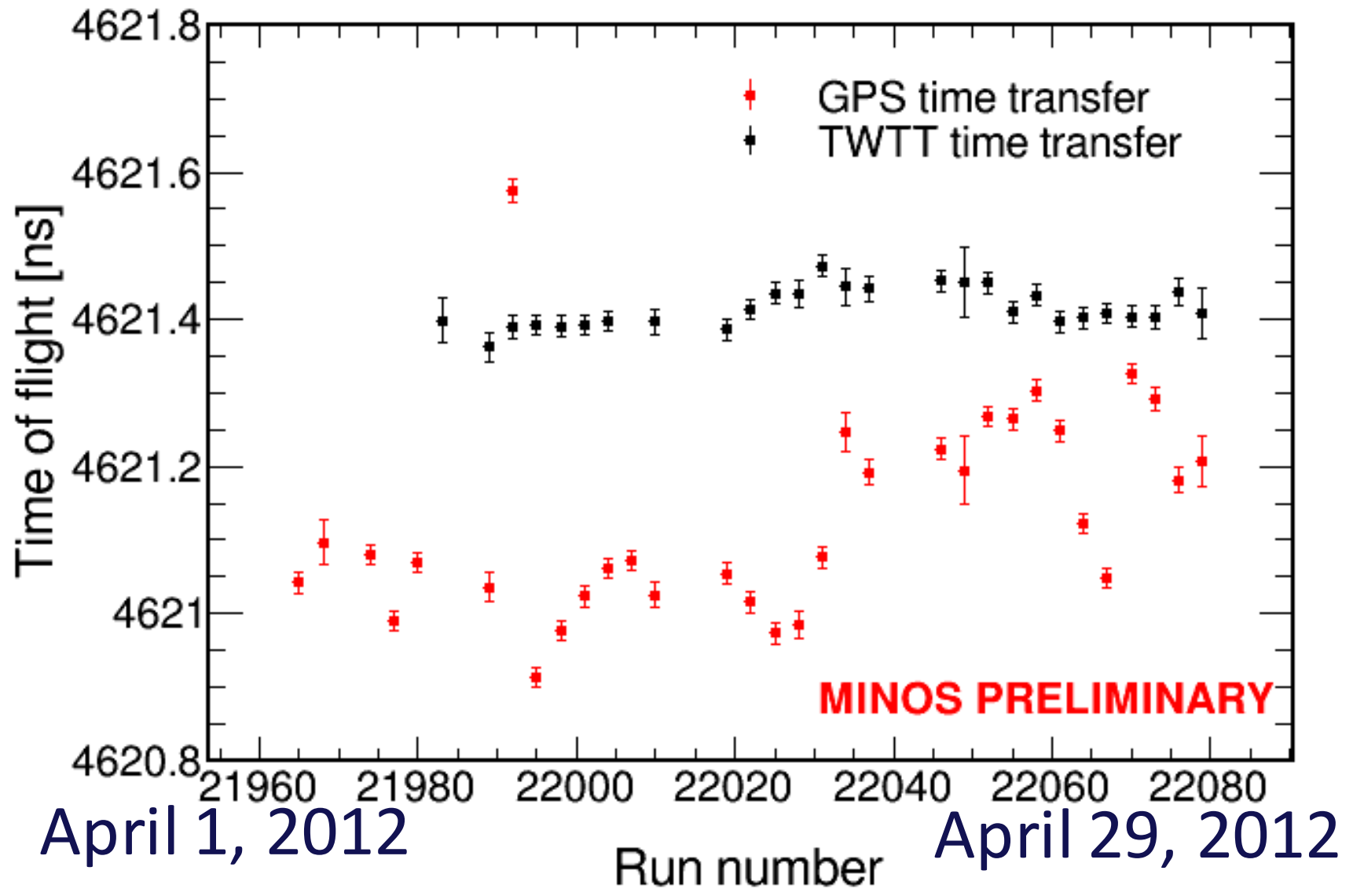

Figure 9. Time of flight between MI60 and Near Detector, using both GPS and fiber TWTT time transfer

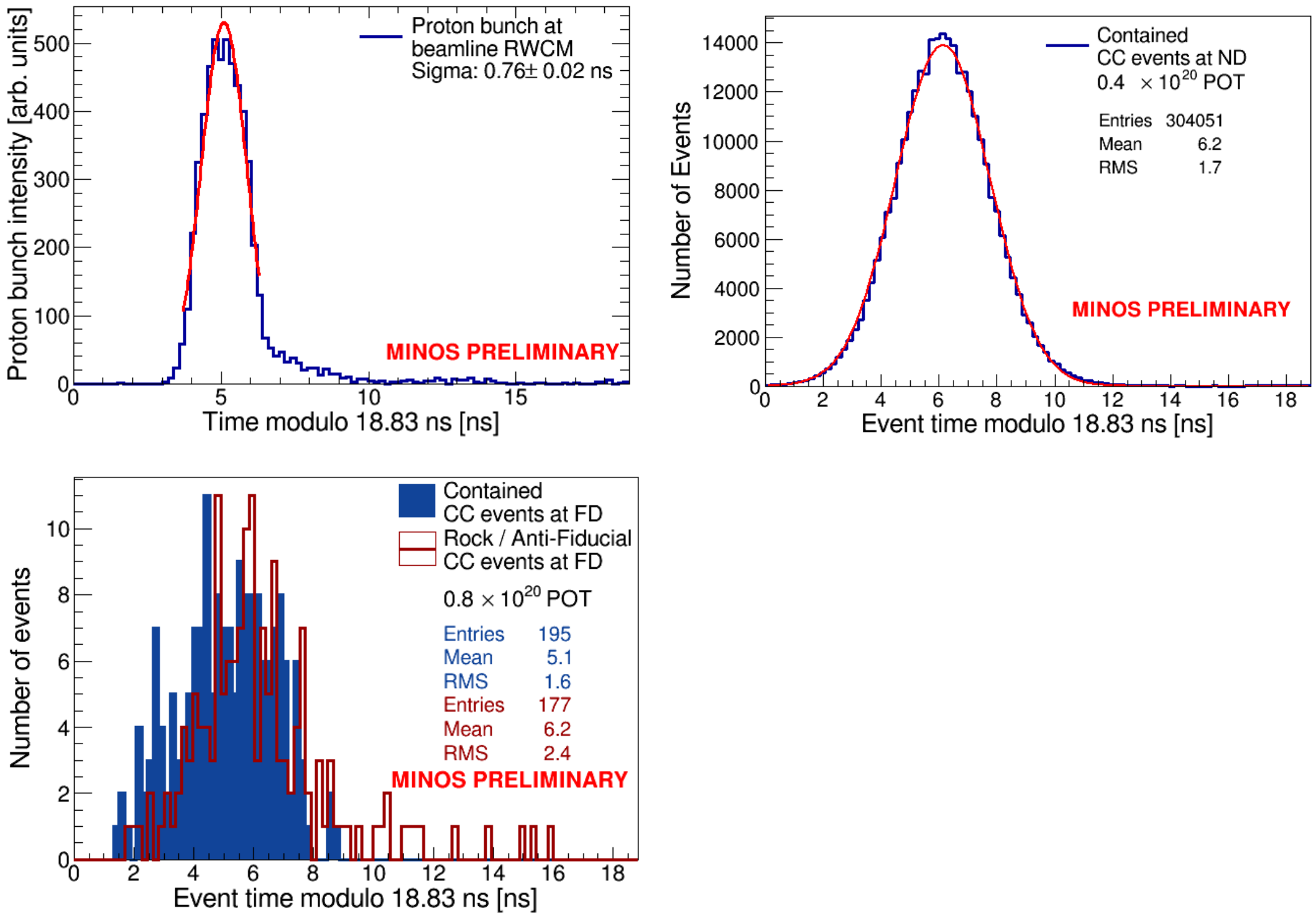

Figure 10. Bunch shape at MI60 RWCM (top left), ND (top right) and FD (bottom). The tail on the RWCM is an artifact caused by an impedance mismatch between the detector and the cable. In the FD plot, the blue histogram shows neutrinos that interacted in the fiducial region away from the edge. The open plot includes "rock" events where the neutrino interacts in the rock surrounding the cavern, and produces a muon that interacts in the detector. These events are expected to arrive on average 1ns later because of the extra path length traveled by the muon.

### *C. Far Detector bunch shape*

Content that the Near Detector data is consistent with expectations, we proceed to the Far Detector. Before we look at the absolute value of the time of flight, we plot the time of flight modulo the MI bunch spacing (18.83 ns). If the time transfer is working and has low jitter, we should see the bunch shape in the distribution. Figure 10 shows the bunch shape at the MI60 RWCM, the ND and the FD. In the Far Detector, we separate the events into two samples – "contained" events (blue histogram) which interact in the fiducial region of the detector away from the edge, and Rock and Anti-Fiducial events (open histogram). Rock events occur when a neutrino interacts in the rock surrounding the detector cavern, producing a muon which then enters the detector. These events usually arrive later than contained events, as the neutrino produces a muon at an angle, so there's a trigonometric increase to the muon's path length. Anti-Fiducial events are neutrino interactions in the very edge of the detector, which are indistinguishable from the incoming muon in a rock event. The net prediction of the detector Monte Carlo simulation is that RAF events should arrive on average 1.0 ns later than contained events, and the distribution should acquire an extra 1.4 ns RMS. The observations in Figure 10 are in good agreement with this prediction.

The appearance of the bunch shape in the FD demonstrates that the whole apparatus is functioning correctly, and also that there is not a large anomalous variation in neutrino velocity as a function of energy.

### *D. The neutrino velocity*

Combining the contained and RAF samples, we measure the time of flight between MI60 and FD as 2,453,935.0 ± 0.1 ns, where we are only considering statistical errors (see Figure 11). Subtracting the earlier measured time of flight between MI60 and ND of 4621.1 ns (using the GPS time transfer value for consistency) we obtain the time of flight between Near and Far Detectors as 2,449,313.9 ± 0.1 ns, where again, only statistical errors are quoted. From [13], the distance between the front face of the Near and Far detectors, including the Sagnac correction, is 2,449,316.3 ± 2.3 ns at the speed of light, where almost all the uncertainty comes from the inertial survey of the Far Detector location. Combining these, together with the other sources of systematic error listed in Table 1, yields a value for the difference in arrival time of the neutrino and the light speed prediction of $\delta = -2.4 \pm 0.1\ (\text{stat.}) \pm\ 2.6\ \text{ns}\ (\text{syst.})$ and so for the neutrino velocity $v$ we have $(v/c - 1) =\ (1.0\ \pm\ 1.1) \times\ 10^{-6}$.

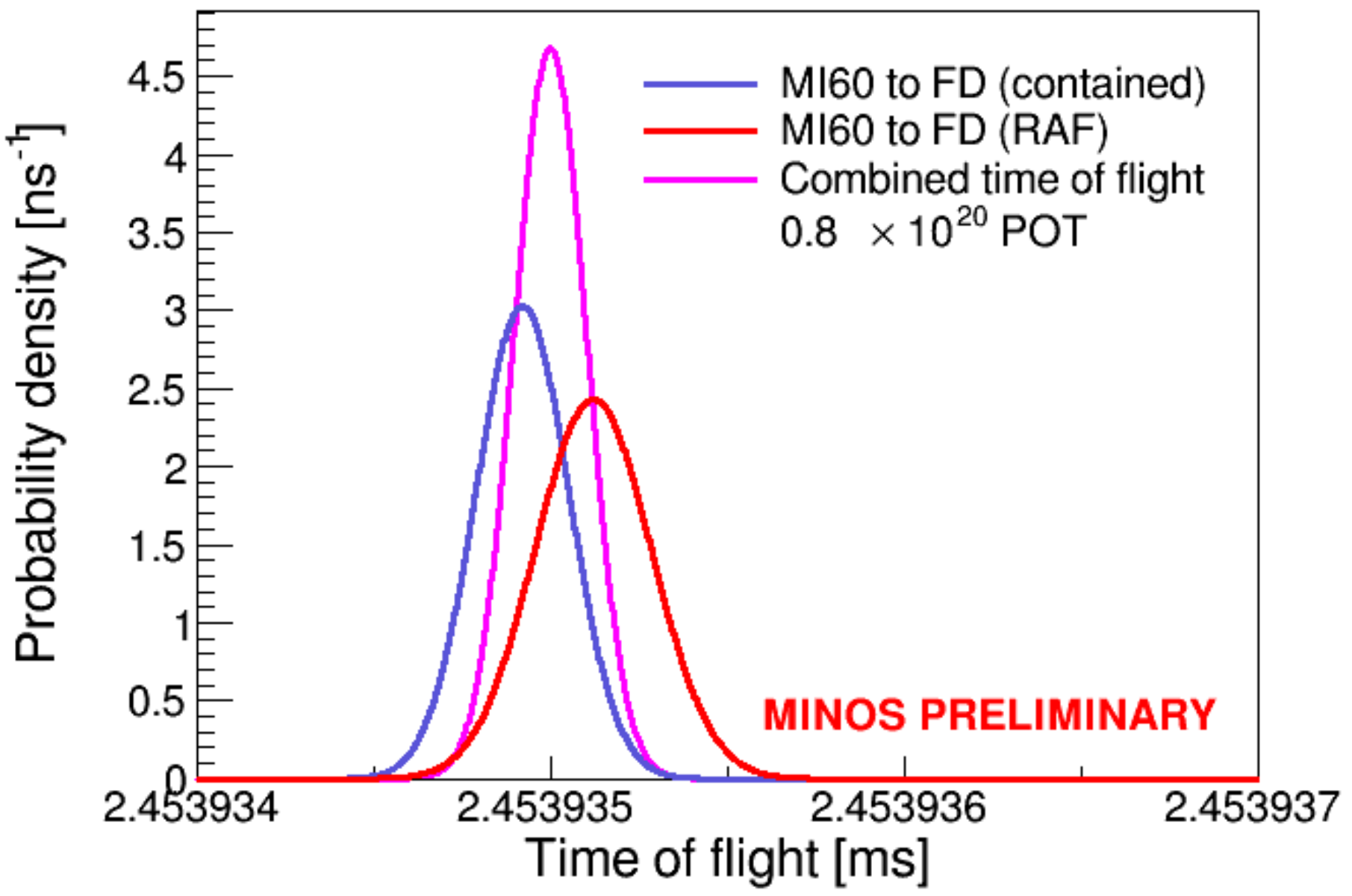

Figure 11. Time of flight between MI60 and Far Detector, showing statistical uncertainties only.

Table I. Systematic uncertainties.

| **Systematic uncertainty** | **Value** |
|---|---|
| Inertial survey at FD | 2.3 ns |
| Relative ND-FD latency | 1.0 ns |
| FD TWTT between surface and underground | 0.6 ns |
| GPS time transfer accuracy | 0.5 ns |
| **TOTAL** | **2.6 ns** |

## V. CONCLUSIONS AND FUTURE SCOPE

We have demonstrated that neutrinos travel at a speed consistent with the speed of light, with an uncertainty which is completely bound by systematics, and dominated by the 2.3 ns uncertainty in the position of the far detector. The total systematic error on the timing portion of the result lies between 1.3 and 1.5 ns.

With further effort, completing the closure of the Auxiliary Detector measurement and having more and better data from the traveling GPS systems, we should be able to reduce the timing error to under a nanosecond. Implementing a conventional optical survey network in the Soudan mine shaft has the potential to reduce the systematic error due to the baseline to under a nanosecond, ultimately measuring the neutrino velocity to a few parts in $10^7$.

When the accelerator complex starts operating again in 2013, the neutrino beam will have been retuned to operate at a higher energy. More neutrinos are produced at these higher energies, and fewer of them will oscillate to a different flavor. This should increase the expected number of $\nu_\mu$ charged-current events at the Far Detector to about 30 per day, yielding a 300 ps statistical precision on the time of flight of the neutrino per day.

## ACKNOWLEDGEMENTS

This work was supported by the US DOE, the UK STFC, the US NSF, the State and University of Minnesota, the University of Athens in Greece, and Brazil’s FAPESP and CNPq. We are grateful to the Minnesota Department of Natural Resources, the crew of Soudan Underground Laboratory, and the staff of Fermilab, for their contributions to this effort.